\documentstyle[10pt]{article}

\begin{document}

\title{\bf Local induction approximation in the theory \\ of superfluid 
turbulence.}
\author{M. V. Nedoboiko \\
Institute of Thermophysics, Novosibirsk, 630090, Russia }
\date{}
\maketitle

\begin{abstract}
The local induction approximation (LIA) of the Biot-Savart law is often used
for numerical and analytical investigations of vortex dynamics (in 
particular in the theory of superfluid turbulence). In this paper,
using renormalization group (RG) methods, some features of the LIA is considered.
The exact statistical solution of the LIA equation is presented.
The problem of "marginal" terms, appearing at the Wilson's approach to the
RG-procedure, is concerned.

\end{abstract}

\section{Introduction.}

It is accepted that chaotic vortex structures, so called 
vortex tangles (VT),
appear in  volume of  superfluid helium under 
the particular conditions. This objects are very interesting because
it is well known that hydrodynamic and thermodynamic 
properties of HeII
in many respects depend on this structures 
\cite{NF95}, \cite{DonBook}.
Obviously, that only statistical consideration is 
possible for the VT
evolution. The vortex tangles consist of separate vortex loops, 
which evolve under
the vortex dynamics laws (the nonlocal Biot-Savart law plus 
interaction of vorticies with the normal component of HeII).
Besides, these loops can merge and break up during temporal evolution. 
These two
mechanisms are not independent, there is their mutual influence.
In general, any analytical investigation of the VT evolution, based on 
consideration of the exact vortex dynamical equations is the
extremely formidable task.
This connected with the absence, at this moment, of any analytical methods 
for solving  nonlinear, nonlocal (as the Biot-Savart), stochastical equations.
Practically all of the known works on this theme represent various numerical
(computer) models to explore properties of VT. In turn, the numerical calculations
also have large difficults. Because the Biot-Savart interaction is the nonlocal 
one,
to determ motion of a single vortex point, it is necessary to take into 
account the influence  
of all entire vortices to this point. Thus, it  reqires unreal computer 
resources and enormous time of the calculations. 

In the initial numerical works  the local induction approximation (LIA) of the
Biot-Savart law was used \cite{Schwarz78}, \cite{Schwarz88}.
 Some analytical advance also connected with 
this method \cite{Schwarz78}. Since then the vast list of numerical works has 
been appeared. The authors obtained various results, using  the Biot-Savart 
law as well as the local induction approximation. However, to this day the 
precision and the validity of the LIA is the objects of discussions. 
  
The processes of reconnections of vortex lines even greater complicate the 
consideration of VT. In this paper we will not concern this problem, and
only expose some properties of the LIA and, in general,  RNG-based solutions
of dynamical tasks.

\section{Dynamical equations.}

To describe the dynamics of a vortex filament we use the 
equation derived in the papers \cite{ Schwarz78}, \cite{Schwarz88}. 
To introduce stochastic evolution we use the
Langevin approach, i.e. we add into the equation   
a random force term. That is the usual way to study turbulent-like 
phenomena. 
As the result, in the local
induction approximation the equation of motion of a
quantized vortex filament in HeII takes the form:

\begin{equation}
\frac{d\vec{s}(\xi ,t)}{dt}=\beta \vec{s^{\prime }}\times \vec{s^{\prime
\prime }}+\nu \vec{s^{\prime \prime }}+\vec{f}(\xi ,t)  \label{initial}
\end{equation}

Here $\vec{s}(\xi ,t)$ is a radius vector of a point of a line 
labeled by the
variable $\xi$, $\vec{s^{\prime }}$ is the derivative on the 
parameter $\xi$; $t$ is time; the quantity $\beta $ is 
the coefficient of nonlinearity
$\beta ~=~\frac \kappa {4\pi }\log \frac R{r_0}$, with the
circulation $\kappa $ and the cutting parameters $R$ 
(the external size, i.e. the averaged radius of
curvature) and $r_0$ (the vortex core size).
The coefficient of dissipation $\nu $ appears in the eq.(1) when 
external counterflow is absent \cite{Schwarz78}, \cite{Schwarz88}. 
The additional term $\vec{f}(\xi ,t)$  models the external random
disturbances such as the "white noise". 
The concrete formulation of $\vec{f}$
will be defined later. Let's note that the eq.(\ref{initial}) 
is valid only if
the normalization of the parameter $\xi$ is the arclength.
This is the additional approximation if to concern the entire consideration
of vortex dynamics, since the overall length 
varies with time. But the aim is only to expose some characteristic features
of the eq.(\ref{initial}). 

In the one-dimensional Fourier-representation 
the equation (\ref{initial}) takes the
form:
\begin{equation}
i\omega s_0^\alpha = \Gamma _{120}^{\alpha \beta \gamma }s_1^\beta
s_2^\gamma -\nu k^2s_0^\alpha +f_0^\alpha ,  \label{initial_f}
\end{equation}
where the operator
\begin{equation}
\Gamma _{120}^{\alpha \beta \gamma }\equiv i\beta \int \int \int \int
\,dk_1\,dk_2\,d\omega _1\,d\omega _2k_1k_2^2\epsilon ^{\alpha \beta \gamma
}\delta (k_1+k_2-k_0)\delta (\omega _1+\omega _2-\omega _0),  \label{Gamma}
\end{equation}
and $s_{k,\omega }$ is  amplitudes (in general, complex) of the Fourier 
harmonics of the function 
$\vec{s}(\xi, t)$, the latin indices denote 
the arguments, i.e. $s_i\equiv
s(k_i,\omega _i)$; the Greek superscripts means the spatial 
components; 
$\epsilon ^{\alpha \beta \gamma }$ is the unit antisymmetric 
tensor. It is
supposed, that the correlator of the random disturbances 
looks like:
\begin{equation}
<\vec{f}(k_1,\omega _1)\vec{f}(k_2,\omega _2)>=Dk^{-y}\delta (k_1+k_2)\delta
(\omega _1+\omega _2)  \label{noise}
\end{equation}

\section{Calculations with the RG method}

Our application of the RG-procedure in principle will be similar 
to the usage of the one in the turbulence theory.  
The details can be found for example in \cite{McComb}, \cite{Foster77},
\cite{Yakhot}.

Let's briefly describe the fundamental principles of the 
RG-method for VT. As usual, the RG-procedure consists of two stages. 
At the first stage we eliminate high (or low) harmonics from the 
consideration. Let us separate the modes 
$s_{k,\omega }$ on low and high harmonicses: 
$s_{k,\omega }^{<}$ and $s_{k,\omega }^{>}$ . 
Here $s_{k,\omega }^{>}=s_{k,\omega }$ at $k\in
(\Lambda e^{-l},\Lambda )$; $s_{k,\omega }^{<}=s_{k,\omega }$ at 
$k\in (K,\Lambda e^{-l})$ and $s^{>},s^{<}\equiv 0$ in the other cases. Thus:
\begin{equation} \label{divide}
s_{k,\omega }=s_{k,\omega }^{>}+s_{k,\omega }^{<}  \label{s_sum}
\end{equation}
The quantity $K$ is the parameter of the infrared cutting.
Since $s_{k=0}=0$ because of the closure of a vortex line, 
the value of $K=2\pi /L$ is defined by
the first non-zero harmonic at the Fourier transforming of 
$\vec{s}$; $L$ denotes the overall length of the vortex filament. 
The parameter of the ultraviolet cutting
$\Lambda $ is defined by the vortex core size $r_0$, 
and in this case $\Lambda
\rightarrow \infty $; the quantity $l$ is 
some positive number, determing a boundary of the modes separation.

Assume that we consider the evolution of any hamonic $s_{k,\omega}$ 
accordingly to
the eq.(2) and $s_{k,\omega}$ is in the interval of low (for example) harmonics.
The idea of the RG method (such as the Wilson's formulation) is not to consider the
interaction of $s_{k,\omega}$ with each of harmonics in the other interval
(the interval of high harmonics in this case). Instead of this, the RG method 
suggests to take into account the ensemble averaged influence of the whole 
high region. This averaging practically is executed on realizations   of 
$\vec{f}$. The interaction between harmonics in own 
(in this case low harmonics) interval is considered in the usual way.

Concretely for a vortex loop this first stage of the RG-procedure is 
realized in the following way. We substitute the rel.(\ref{s_sum}) into 
the eq.(\ref{initial_f}) and then averaging  $s^{>}$ over the 
ensemble of $\vec{f}$. 
As the result we obtain a new, modified equation instead of 
the eq.(\ref{initial_f}). 
Note that the value 
$\Lambda$ is transformed as $\Lambda\rightarrow \tilde{\Lambda}=
\Lambda e^{-l}$. This new equation doesn't describe behaviour of VT on 
small spatial scales --- the 
region of wave vectors $(\Lambda e^{-l}\div \Lambda)$ becomes unavailable. 
Such procedure is similar to the Kadanoff transformation \cite{Ma80}.
If this new equation has the same form (perhaps approximately) as the original 
one  and only it's parameters are changed, 
the second stage of the RG-procedure can be done.

On the second stage  we 
reconvert the quantity $\tilde{\Lambda}$ to the original value, 
i.e. we are returning to the description on the initial scale. 
In the other words, we are performing the transformation 
$\tilde{\Lambda}\rightarrow \Lambda =\tilde{\Lambda}e^{l}$.

Let's analyse some possible transformations of the parameters at the first 
stage. For example, in the turbulence theory the RG-procedure gives the renormalization
of viscosity. This value has a positive addition after averaging high
modes. In the other words, the influence of small spatial
scales (or high harmonics) leads to  increasing the viscosity. 
That, in turn, means that the 
dissipation of energy is increased. The mechanism of that phenomena is obvious.
There are two processes for the evolution of any harmonic. The first is the
thermal dissipation, 
the second is the nonlinear interaction with other harmonics,
or, in the other words, the transfer of the energy of turbulent pulses 
across the spectrum in the 
wave-number space. (We don't take into account the external random 
disturbance $\vec f$).
Averaging  high harmonics lead to increasing the dissipation of low harmonics.
This mean the existence of the transfer of the harmonic amplitudes from 
the low-frequency part of the spectrum to the high-frequency one. 
And just the transfer is the mechanism of the additional dissipation.
Ones again emphasize: the additional dissipation (increasing viscosity)
connected with the influence of the opposite spatial region, in the other
words, the flux across the spectrum is responsible for increasing dissipation.
And vice versa, a change of the viscosity under the averaging, means, in turn,  
the presence of the flux across the spectrum in a considering dynamical system.
For turbulence theory this result is well known and trivial.
Let's assume that we have solved some dynamical task and in an analogous case,
we have obtained a {\it negative} addition to a dissipative coefficient. It must mean
that the transfer across the wave-number spectrum has the opposite direction
(i.e. from high modes to low ones). The zero
addition must mean the absence of the transfer.

Now let's study the eq.(1)-eq.(2) in this context. The substitution of the
rel.(\ref{divide}) into the eq.(2), gives the result:
\begin{equation}\label{div_init}
i\omega s_0^{\alpha <} = \Gamma _{120}^{\alpha \beta \gamma }\:\{
s_1^{\beta <} s_2^{\gamma <}+
<s_1^{\beta <} s_2^{\gamma >}>+
<s_1^{\beta >} s_2^{\gamma <}>+
<s_1^{\beta >} s_2^{\gamma >}>\}
-\nu k^2s_0^{\alpha <}+f_0^{\alpha <}
\end{equation} 
The brackets $<...>$ mean the averaging of only high harmonics. To do this
averaging,  turn to the functional integral formalism \cite{Jensen}.
In this approach, any correlator (for example the fourth term in 
the right hand side of the eq.(\ref{div_init})) is represented as a product 
of the functional derivatives (below denoted as $\frac{\delta}{i\delta\eta}$ or 
$\frac{\delta}{i\delta\hat\eta}$) on auxiliary fields ($\eta\: ,\hat\eta$)
from the characteristic              
(generating) functional of a dynamical system. Not going into details 
\cite{Jensen}, we write the expression to average the fourth therm:
$$
<s_1^{\beta >} s_2^{\gamma >}> = \frac{\delta}{i\delta\eta^{\beta}(1)}
\frac{\delta}{i\delta\eta^{\gamma}(2)}\:\Bigm \{
e^{\int d0\:\Gamma^{\alpha\beta\gamma}_{120}\{ \frac{\delta}{i\delta\hat
\eta^{\alpha}(0)}
\frac{\delta}{i\delta\eta^{\beta}(1)}\frac{\delta}{i\delta\eta^{\gamma}(2)}\}}
\Bigr\}
$$
\begin{equation}\label{cf}
\: e^{\int d1\: i\eta^\alpha (1) G^{(0)\:\alpha\beta}(1) \hat \eta^\beta (-1)- 
1/2\eta^\alpha (1) C^{(0)\:\alpha\beta}(1) \eta^\beta(-1)}\:\Bigr |_{\eta =\hat\eta =0}
\end{equation} 
Here $G^{(0)}(1)$ is the zero-order response function  $G^{(0)\:\alpha
\beta}(1)=
\frac{\delta^{\alpha\beta}}{i\omega _1 +\nu k^2_1}$; $C^{(0)}(1)$ is the 
zero-order  correlation 
function  $C^{(0)\:\alpha\beta}(1)=\frac{D\delta^{\alpha\beta}}{(\omega _1^2+\nu^2 k_1^4)
k^y}$ (to clarify $\frac{D}{k^y}$ see the rel.(4)).
 
It is well known that the Taylor-series of the first exponent
gives usual Feynman's diagram series. 
In our case this takes the form:
$$
<s_1^{\beta >} s_2^{\gamma >}> = \frac{\delta}{i\delta\eta^{\beta}(1)}
\frac{\delta}{i\delta\eta^{\gamma}(2)}\:\Bigm \{1+\Gamma_{120}^{\alpha\beta
\gamma}\{\frac{\delta}{i\delta\hat
\eta^{\alpha}(0)}
\frac{\delta}{i\delta\eta^{\beta}(1)}\frac{\delta}{i\delta\eta^{\gamma}(2)} \}+
\Gamma^{\alpha\beta\gamma}_{120}\{...\}\Gamma_{120}^{\delta\theta\zeta}\{...\}
\: +\: .\: .\: .\Bigr \} 
$$
\begin{equation}\label{cf1} 
e^{i\eta  G^{(0)} \hat \eta  - 
1/2\eta C^{(0)} \eta}\:\Bigr |_{\eta =\hat\eta =0}
\end{equation}
The essential integrations in the last formulae are assumed.
Thus, the result for $<\vec s\; \vec s>$ represents the infinite series of terms, each of that
is a product of $C^{(0)}$, $G^{(0)}$ functions (corresponding to the lines, 
if diagrams to consider) and essential integrations.  Every line 
$C^{(0)}$ or $G^{(0)}$ is the result
of coupling the derivatives: $\{\frac{\delta}{i\delta \eta}
\frac{\delta}{i\delta \eta}\}$  or $\{\frac{\delta}{i\delta \eta}
\frac{\delta}{i\delta \hat\eta}\}$ accordingly. Obviously only terms,
containing an even number of the derivatives is not equal to zero. 
Otherwise, $\eta$ or $\hat\eta$ remains as
the efficient in the product and the term equal to zero because 
$\eta =\hat\eta =0$. Besides
if $C^{(0)\:\alpha\beta}$ or $G^{(0)\:\alpha\beta} (\alpha \ne\beta)$ is presented in a
product, this term is equal to zero, because $\delta^{\alpha\beta}$ is
presented in the definition of $G^{(0)\:\alpha\beta}$ and $C^{(0)\:\alpha\beta}$. 
It is easy to observe that {\it all} terms of the series for the exact 
correlator $<s^\alpha\; s^\beta>$ ($\alpha\ne\beta$) are
identically equal to zero. Since $\alpha\ne\beta$ in the external 
diferentiation,
in the internal differentiation an {\it odd} number of the derivatives having 
$\alpha$ or $\beta$-components are remained and thus the line $G^{(0)\:
\alpha\beta}$ or $C^{(0)\:\alpha\beta}$ is {\it necessarily} appeared. 
{\it Thus, $<s^\alpha \; s^\beta>$ ($\alpha\ne\beta$) equal to zero and this 
is the exact result.} 

Similar argumentation leads to $<s^\alpha>$ is equal to zero. Hence all additional
terms in the eq.(\ref{div_init}) exactly equal to zero. The latter means the 
eq.(\ref{div_init}) doesn't change under the averaging. Thus,  the transfer
whithin the spectrum in the LIA is absent. 
And this result is exact.

\section{Discussion and conclusions.}
Thus we obtained that the interaction of the form: $\vec{s^\prime}\times
\vec{s^{\prime\prime}}$ doesn't change the spectral structure of the value $\vec s$.
In the other words, though there is the nonlinear interaction in the system,
harmonics doesn't affect to each other. This is the extremely unexpected result. 
The usual scenario looks quite otherwise. Usually, because nonlinearity is
present, interaction of only two harmonics leads to the appearence of
additional harmonics. They, in turn, interact between each other and with
original ones. As the result, after some time, the full spectrum is presented 
in the system.  In our case in the system remain only originally excited
modes and their amplitudes are not changed. At the same time, their phases
($s_{k\omega}$ is a complex number)
must be changed.
Otherwise it should  be full correspondense to a linear equation. However,
it is well known the solution (that used the LIA) of the task  about the 
decay of kink \cite {Buttke}, that is not possible in a linear case.

Obviously, it is possible to invent an infinite number of similar equations,
not having a flow in the wave-number space.  Perhaps, all of them present
pure abstract interest. Nevertheless, the single practical note can be done.
As mentioned above, the LIA is often used to approximate the full
Biot-Savart law in numerical models of superfluid turbulence. The question
about the validity of this approach was dicussed repeatedly and the exact 
opinion is absent. Let's consider this problem in the context of this work.
Transfer within the spectrum is  certanly presented in the entire Biot-Savart 
law, at that, probably, the nonlocal transfer. (Real vortex tangle is 
described by the Biot-Savart law). Besides, the transfer within the 
spectrum is the very important feature of nonlinear dynamical systems. 
For example, whether is it possible to
imagine  any approach to a turbulence theory, where the transfer of
turbulent pulses from large scales to small ones is absent...
But just the same picture  takes place, when the LIA is used for numerical
or analytical considerations of real VT.

The last remark concerns to the usage of RG-procedures for dynamical problems.
At this moment this question is not fully clear. Usual way  emploing
this method based on a perturbation theory. As a result, at the Wilson's
formulation of  RG, an infinite series of additional terms is appeared
in the renormalized equations. So-called "marginal" terms represent, at
this case, the main problem. The total value of this additions is 
uncontrolled when the Wilson's RG-procedure is carried out. The detal 
consideration of this question can be found in the paper \cite{Eyink}. 

In principle, we can expand $s^>$, $s^<$ to a perturbation series.
In the other words, to employ the usual approach to the RG-procedure. 
If we have done this, the following result should be obtained.
There is  transfering  amlitudes
of harmonics across the spectrum is existed. And it is interesting to note, 
that the flux of harmonics is directed from small spatial scales to large 
ones, i.e. to the opposite side, in comparison with the turbulence theory. 
It is very unexpected result for superfluid turbulence, if it were right.
In real, the LIA does not describe the flux of amplitudes of harmonics within
the spectrum.  

Returning to the RG-method and comparing perturbation theory with the presented 
above exact solution, one can conclude that the marginal additions 
are really uncontrolled, even if $\epsilon$-expansion to use.

\bigskip
The work was carried out under
support of INTAS (grant N 2001-0618).

\end{document}